\documentclass[submission, Proceedings]{SciPost}
\usepackage{graphicx}
\usepackage{float}
\usepackage{caption}
\usepackage{subcaption}

\hypersetup{
    colorlinks,
    linkcolor={red!50!black},
    citecolor={blue!50!black},
    urlcolor={blue!80!black}
}

\usepackage[bitstream-charter]{mathdesign}
\DeclareSymbolFont{usualmathcal}{OMS}{cmsy}{m}{n}
\DeclareSymbolFontAlphabet{\mathcal}{usualmathcal}

\begin{document}
	
\begin{center}{\Large \textbf{
Measurement of Transverse Spin Dependent Azimuthal Correlations of Charged Pion(s) in  $p^{\uparrow} p$ Collisions at $\sqrt s  = 200$ GeV at STAR\\
}}\end{center}

\begin{center}
Babu R. Pokhrel\textsuperscript{1,*}, for the STAR Collaboration
\end{center}
\begin{center}
{\bf 1} Temple University, Philadelphia, USA
\\
* babu.pokhrel@temple.edu
\end{center}

\begin{center}
\today
\end{center}

\definecolor{palegray}{gray}{0.95}
\begin{center}
\colorbox{palegray}{
  \begin{tabular}{rr}
  \begin{minipage}{0.1\textwidth}
    \includegraphics[width=22mm]{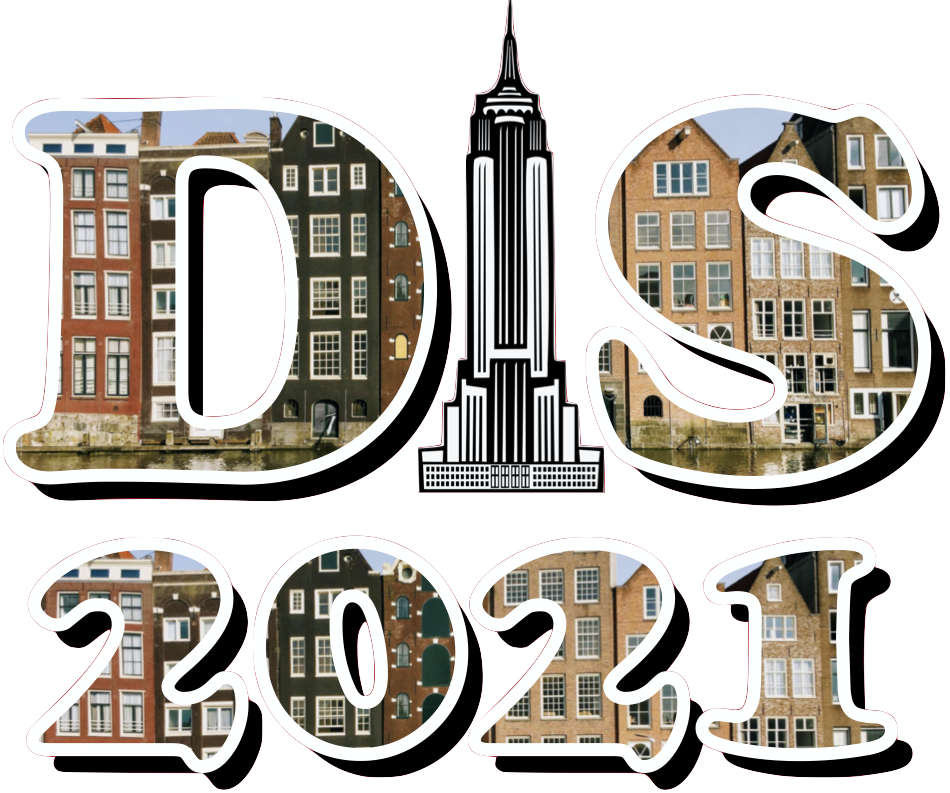}
  \end{minipage}
  &
  \begin{minipage}{0.75\textwidth}
    \begin{center}
    {\it Proceedings for the XXVIII International Workshop\\ on Deep-Inelastic Scattering and
Related Subjects,}\\
    {\it Stony Brook University, New York, USA, 12-16 April 2021} \\
    \doi{10.21468/SciPostPhysProc.?}\\
    \end{center}
  \end{minipage}
\end{tabular}
}
\end{center}

\newlength{\halfwidth}
\setlength{\halfwidth}{\dimexpr 0.5\textwidth-\tabcolsep}

\section*{Abstract}
{\bf At the leading twist, the transversity distribution function, $h^{q}_{1}(x)$, where $x$ is the longitudinal momentum fraction of the proton carried by quark $q$, encodes the transverse spin structure of the nucleon. Extraction of it is difficult because of its chiral-odd nature. In transversely polarized proton-proton collisions ($p^\uparrow p$), $h_{1}^{q}(x)$ can be coupled with another chiral-odd partner, a spin-dependent fragmentation function (FF). The resulting asymmetries in hadron(s) azimuthal correlations directly probe  $h_{1}^{q}(x)$.
We report the measurement of correlation asymmetries for charged pion(s) in $p^\uparrow p$, through the Collins and the Interference FF channel.  
}

\section{Introduction}
At the leading twist, the nucleon structure is fully described by three Parton Distribution Functions (PDFs): the unpolarized PDF,  $f_{1}(x)$, the helicity PDF, $g_{1}(x)$,  and the transversity PDF, $h^{q}_{1}(x)$, where  $x$ is the nucleon momentum fraction carried by partons. Although, $f_1(x) $ and $g_{1}(x)$ are reasonably well constrained by experimental data \cite{EurPhys,PhysRevD.80.034030}, the knowledge of $h_1^{q}(x)$ is limited to the semi inclusive deep inelastic scattering (SIDIS) and $e^+e^-$ data \cite{transversity}. This is because $h_1^{q}(x)$ is a chiral-odd object and it needs to be coupled with another chiral-odd partner to form a chiral-even cross section that is experimentally observable.

In polarized proton-proton collisions ($p^\uparrow p$), $h_{1}^{q}(x)$ can be coupled with chiral-odd spin-dependent fragmentation functions (FFs). Selecting inclusive charged hadrons within jets, collimated sprays of particles produced by fragmentation and hadronization of partons in high energy collisions, involves the Collins FF, whereas selecting oppositely charged di-hadron pairs in the final state involves the interference FF (IFF). In both channels, the coupling of $h_{1}^{q}(x)$ with the respective FF results in experimentally measurable azimuthal correlation asymmetry, $A_{UT}$, which is sensitive to $h_{1}^{q}(x)$.  

\section{Experiment and Dataset}
The Relativistic Heavy-Ion Collider (RHIC) at Brookhaven National Laboratory (BNL) is capable of colliding bunched beams of polarized protons up to a center-of-mass energy ($\sqrt{s}$) of  510 GeV. The Solenoidal Tracker At RHIC (STAR) is one of the major experiments, where the Time Projection Chamber (TPC) is the main detector that provides particle tracking and identification in the mid-pseudorapidity region ($-1\textless \eta \textless 1$) and over the whole $2\pi$ range in azimuthal angle \cite{ackermann2003star}. The time-of-flight detector (TOF) \cite{tof}, with a similar coverage as the TPC, improves the STAR's PID capability. The barrel electromagnetic calorimeter (BEMC) provides event triggering based on the energy deposited in its towers.

STAR firstly observed the IFF asymmetry based on 2006 $p^\uparrow p$ data at $\sqrt{s}=200$ GeV \cite{adamczyk2015observation}, followed by the 2011 data at $\sqrt{s}=500$ GeV \cite{adamczyk2018transverse}, and the Collins asymmetry based on 2011 data at $\sqrt{s}=500$ GeV \cite{STARCollins2018}.
STAR collected additional $p^\uparrow p$ data at $\sqrt{s}=$ 200 GeV in 2012 and 2015. These datasets correspond to the integrated luminosities, $L$, of $\sim 14\ \mathrm{pb^{-1}}$ and $\sim 52\ \mathrm{pb^{-1}}$, respectively, with the average beam polarization of $\sim 58\%$.  They provide the most precise measurements of the Collins and the IFF asymmetries in $p^\uparrow p$ at $\sqrt{s}=200$ GeV to date, especially at quark momentum fractions $0.1\textless x\textless 0.4$. The combined 2012 and 2015 dataset is used for the Collins analysis and only 2015 dataset is used for the IFF analysis. 

\section{Results}
The Collins and the IFF asymmetries for charged pion(s) are extracted using the cross-ratio formula \cite{OHLSEN197341},
	\begin{equation}
A_{UT}\cdot sin(\phi) =\frac{1}{P}\cdot \frac{\sqrt{N^\uparrow_{1,\alpha} N^\downarrow_{1,\beta} } - \sqrt{N^\downarrow_{1, \alpha} N^\uparrow_{1,\beta} }}{\sqrt{N^\uparrow_{1,\alpha} N^\downarrow_{1, \beta} } + \sqrt{N^\downarrow_{1,\alpha} N^\uparrow_{1,\beta}}}
\end{equation}
where, $N^{\uparrow(\downarrow)}$ is the number of $\pi^{\pm}$ within jets (Collins channel) or exclusive $\pi^+\pi^-$ pairs (IFF channel) when the beam polarization is $\uparrow(\downarrow)$, in the respective detector halves, $\alpha$ and $\beta$. $P$ is the average beam polarization. The azimuthal angle definitions and asymmetry extraction approach for the IFF and the Collins channels are based on the STAR publications \cite{adamczyk2015observation} and \cite{STARCollins2018}, respectively. 
The mechanism of producing azimuthal correlations and its extraction from a theoretical point of view can be found in \cite{Bacchetta:2004it}.

High-quality tracks are selected by applying several quality cuts and charged pions are identified by measuring their ionization energy loss, $\langle dE/dx \rangle$. For both channels, pions are selected by requiring a cut on the number of standard deviations of measured $\langle dE/dx\rangle$ from the expected pion energy loss,  $-1\textless n\sigma_{\pi}\textless 2$. Furthermore, we find that the TOF enhances the particle identification (PID) in the momentum region where the TPC $dE/dx$ between particle species overlaps. The Collins analysis utilizes both TPC and TOF information for PID in those regions, whereas the IFF analysis only makes use of the TPC. For both analyses, the average $\pi^\pm$ purity reaches $\sim$90\% in different kinematic regions. However, IFF analysis uses $\pi^+\pi^-$ pairs, whose combined purity is $\sim 80\%$.

To estimate the trigger bias on the measurements, PYTHIA 6 \cite{pythia} events are run through the STAR detector simulation implemented in GEANT 3 \cite{geant} and embedded into zero-bias events.
 The magnitude of the bias is determined by calculating the fraction of quark events at the detector level (GEANT) and at the particle level (PYTHIA) and taking a ratio between them.
 The effect of particle impurity and the trigger bias correction are the two main sources of systematic uncertainties.
 \begin{figure}[H]
 	\centering
 	\begin{subfigure}[t]{0.45\textwidth}
 		\centering
 		\includegraphics[scale=0.4, trim= 0.3cm 0.2cm 0cm 0.5cm, clip]{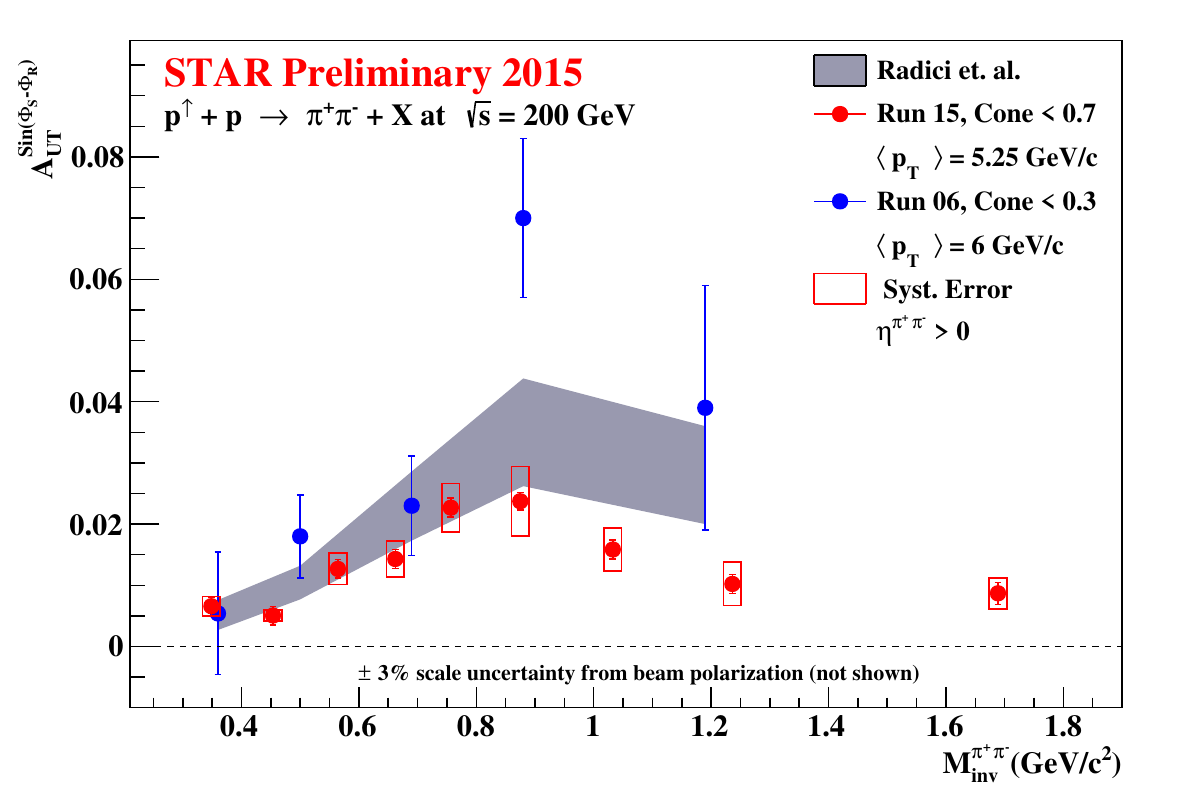}
 		\captionsetup{justification=centering}
 		\caption{$A_{UT}^{sin(\phi_{s}-\phi_{R})}$ vs $M_{inv}^{\pi^+\pi^-}$}
 		\label{iff1}
 	\end{subfigure}
 	\hspace{1cm}
 	\begin{subfigure}[t]{0.45\textwidth}
 		\centering
 		\includegraphics[scale=0.35, trim= 0.1cm 0.2cm 0cm 0.5cm, clip]{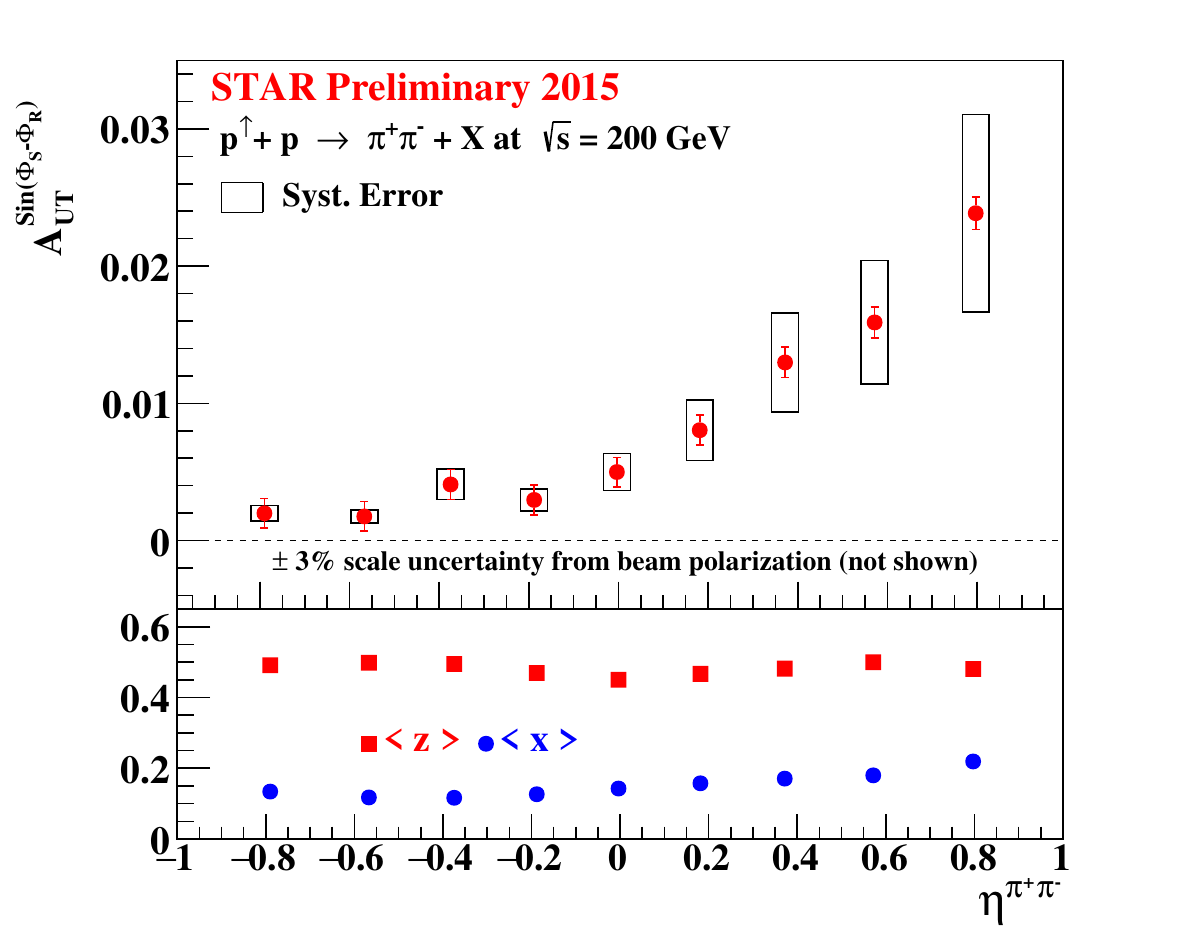}
 		\caption{ $A_{UT}^{sin(\phi_{s}-\phi_{R})}$ vs $\eta^{\pi^+\pi^-}$} 
 		\label{iff2}
 	\end{subfigure}
 	\caption{STAR IFF asymmetries: \ref{iff1}) $A_{UT}^{sin(\phi_{s} -\phi_{R})}$ as a function of invarinat mass of two oppositely-charged pions, $M_{inv}^{\pi^+\pi^-}$, in $\eta^{\pi^+\pi^-} \textgreater 0$ region, compared with the theoretical calculation from \cite{mradici}. The \emph{cone} cut ($\sqrt{(\eta^{\pi^+}-\eta^{\pi^-})^2 + (\phi^{\pi^+}-\phi^{\pi^-})^2}\textless 0.7$ ) ensures that the $\pi^+$ and $\pi^-$ are close enough in $\eta-\phi$ space.
 		\ref{iff2}) $A_{UT}^{sin(\phi_{s} -\phi_{R})}$ as a function of  $\eta^{\pi^+\pi^-}$, integrated over $M_{inv}^{\pi^+\pi^-}$ and $p_{T}^{\pi^+\pi^-}$ (\emph{top panel}). The quark $\langle z \rangle$ and $\langle x \rangle$, in the corresponding $\eta^{\pi^+\pi^-}$ bins, are shown in the \emph{bottom panel}.
 	}
 \end{figure}
\begin{figure}[H]
	\centering
	\includegraphics[scale=0.35, trim= 0.1cm 0.0cm 0cm 0.2cm, clip]{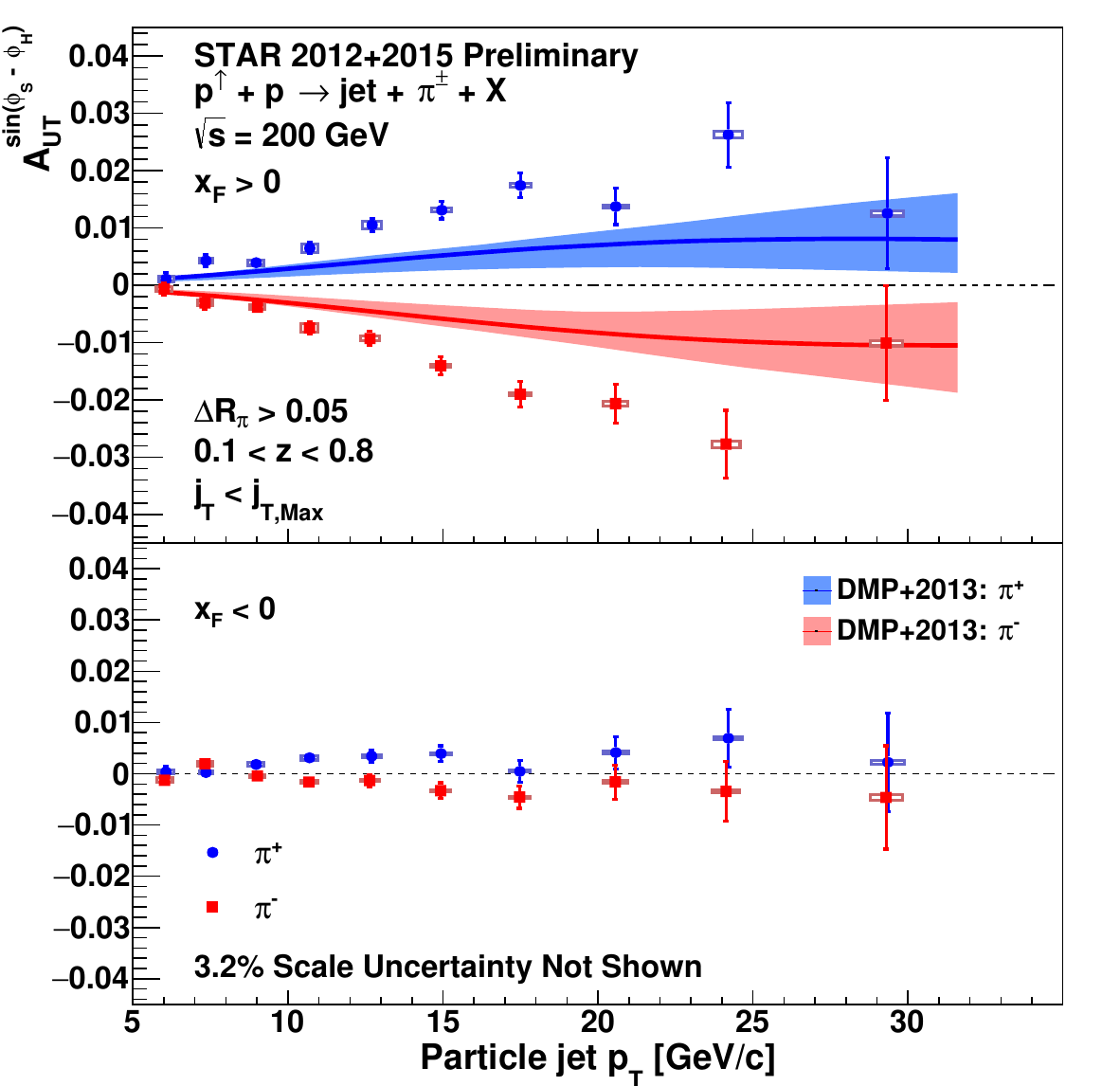}
	\caption{STAR Collins asymmetry: $A_{UT}^{sin(\phi_s -\phi_H)}$ as a function of  the particle-jet $p_T$ in forward ($x_{f}\textgreater 0$) (\textit{top panel}) and backward ($x_f\textless 0$) jet scattering directions (\textit{bottom panel}), for the $\pi^\pm$ within jets, compared with the theoretical calculation from \cite{collinstheory}.} 
	\label{collins}
\end{figure}

 Figure \ref{iff1} depicts preliminary results for the IFF asymmetry, $A_{UT}^{sin(\phi_s -\phi_R)}$, as a function of invariant mass of $\pi^+\pi^-$ pair, $M_{inv}^{\pi^+\pi^-}$, in forward  $\pi^+\pi^-$ pseudorapidity ($\eta^{\pi^+\pi^-}>0$) region. It is integrated over the transverse momentum of the $\pi^+\pi^-$ pair, $p_T^{\pi^+\pi^-}$, in the interval 2.5 to 15 GeV/$c$. The $A_{UT}^{sin(\phi_s-\phi_R)}$ signal is enhanced around $M_{inv}^{\pi^+\pi^-}\sim 0.8$ $\mathrm{GeV}/c^{2}$, which is consistent with the previous STAR measurements \cite{adamczyk2015observation, adamczyk2018transverse} and the theoretical calculation at $\sqrt{s}=200$ GeV \cite{mradici} incorporating SIDIS, $e^+e^-$, and STAR 2006 $p^\uparrow p$ results. This enhancement, close to $\rho-$meson mass ($M_\rho \sim0.775 \ \mathrm{GeV}/c^2$), is expected and consistent with a IFF model calculation \cite{jaffe1998interference}. The corresponding $A_{UT}^{sin(\phi_s -\phi_{R})}$ in the backward pseudorapidity region ($\eta^{\pi^+\pi^-}<0$) is small. 
 
Figure \ref{iff2} shows $A_{UT}^{sin(\phi_s -\phi_R)}$ as a function of $\eta^{\pi^+\pi^-}$, integrated over $M_{inv}^{\pi^+\pi^-}$ and $p_{T}^{\pi^+\pi^-}$ (\textit{upper panel}). The average $x$, fractional proton momentum carried by a quark, and $z$, fractional quark energy carried by the $\pi^+\pi^-$ pair, are estimated from GEANT simulation in the corresponding $\eta^{\pi^+\pi^-}$ bins and shown in the bottom panel. $A_{UT}^{sin(\phi_s-\phi_R)}$ increases linearly with $\eta^{\pi^+\pi^-}$ in the forward region. The small asymmetry signal in the backward $\eta^{\pi^+\pi^-}$ region is mainly due to scattering from a quark at  lower $x$, which is typically associated with the unpolarized beam. A strong correlation between the observed asymmetry and $x$ can be seen, where $x$ ranges from $\sim 0.1$ to $ 0.22$ from backward to forward $\eta^{\pi^+\pi^-}$. However,  $z$ shows no clear dependence, the average of which is $\sim 0.46$. The 2015 IFF results corroborate previous 2006 \cite{adamczyk2015observation} and 2011 \cite{adamczyk2018transverse} results. 

\indent Preliminary results for the Collins asymmetry, $A_{UT}^{sin(\phi_s -\phi_H)}$, as a function of particle jet $p_{T}$ are shown in figure \ref{collins}. A significant positive asymmetry for $\pi^+$ and negative asymmetry for $\pi^-$ is observed in the $x_f \textgreater 0$ region (\textit{upper panel}). Though small, $A_{UT}^{sin(\phi_s -\phi_H)}$ follows a similar charge dependence in the $x_f\textless 0$ region as well (\textit{lower panel}). This charge-dependence is consistent with a theoretical calculation \cite{collinstheory} and the Collins asymmetry in SIDIS \cite{SIDISCollinsJlab}. 
Although the theoretical calculation undershoots data, they both follow a similar trend. This result shows a large asymmetry signal with higher statistical precision than previous STAR Collins analysis \cite{STARCollins2018}. The Collins analysis is also performed for the identified kaon ($K$)  and proton ($p$). It is found that the Collins asymmetry for $K^+$ is about the size of $\pi^+$ within the statistical uncertainties, while $K^-$ and $p(\bar{p})$ asymmetries are consistent to zero.

\section{Conclusion}
\label{sec:conclusion}
STAR has measured charged pion(s) correlation asymmetries through the IFF channel based on 2015 and the Collins channel based on 2012+2015 $p^\uparrow p$ data at $\sqrt{s}=200$ GeV. These datasets cover the $Q^2$ at the order of $\sim 100\ \mathrm{GeV^2}$ at intermediate $x$, which is well within the valance quark region.
The measured IFF asymmetry signal is enhanced around $M_{inv}^{\pi^+\pi^-}\sim 0.8 \ \mathrm{GeV}/c^2$, which is consistent with the theoretical calculation and the previous STAR measurements. 
 A large asymmetry in the forward $\eta^{\pi^+\pi^-}$ region corresponds to higher $x$, where quark transversity is expected to be sizeable, whereas the backward asymmetries are small since the probed low-$x$ quarks are mainly from the unpolarized proton.
The large Collins asymmetry, as a function of particle jet $p_T$, is larger than the theory prediction in the $x_f\textgreater0$ region, but exhibits a similar trend, whereas the asymmetry in the $x_f \textless 0$ region is small. The charge-dependence of $\pi^+(\pi^-)$ asymmetry is consistent with the Collins asymmetry found in SIDIS.
 The statistical precision of these results is largely improved with respect to previous STAR results. The systematic uncertainty includes the effect from the PID and trigger bias, which is well understood in the Collins analysis. However, the large systematic uncertainty in the IFF analysis is dominated by the PID effect, which will be reduced in the near future. These high percision IFF and Collins asymmetriey measurements will help to constrain the valance-quark transversity distributions and test the universality of the mechanism producing such asymmetries in different collision processes: SIDIS, $e^+e^-$, and $p^\uparrow p$.          

\section*{Acknowledgements}
We thank Marco Radici (INFN Pavia, Italy) for providing IFF asymmetry theory curve.
 We thank the RHIC Operations Group and RCF at BNL. This work was supported by the Office of the Nuclear  Physics within the U.S. DOE Office of Science. 

\bibliography{SciPost_LaTeX_Template.bib}
\nolinenumbers

\end{document}